\documentclass[useAMS,usenatbib]{mn2e}

\usepackage{amsmath,amssymb,graphicx}

\newcommand{\kepler}{\textit{Kepler}}

\title[Kepler's missing planets]{Kepler's missing planets}
\author[Steffen \textit{et al.}]{
Jason H. Steffen$^{1,2}$
\\
$^{1}$Northwestern University, 2131 Tech Drive, Evanston, IL 60208\\
$^{2}$CIERA Fellow\\
}

\begin{document}


\pagerange{\pageref{firstpage}--\pageref{lastpage}} 

\maketitle

\label{firstpage}

\begin{abstract}
We investigate the distributions of the orbital period ratios of adjacent planets in high multiplicity \kepler\ systems (four or more planets) and low multiplicity systems (two planets).  Modeling the low multiplicity sample as essentially equivalent to the high multiplicity sample, but with unobserved intermediate planets, we find some evidence for an excess of planet pairs between the 2:1 and 3:1 Mean Motion Resonances in the low multiplicity sample.  This possible excess may be the result of strong dynamical interactions near these or other resonances or it may be a byproduct of other evolutionary events or processes such as planetary collisions.  Three planet systems show a significant excess of planets near the 2:1 Mean Motion Resonance that is not as prominent in either of the other samples.  This observation may imply a correlation between strong dynamical interactions and observed planet number---perhaps a relationship between resonance pairs and the inclinations or orbital periods of additional planets.  The period ratio distributions can also be used to identify targets to search for missing planets in the each of the samples, the presence or absence of which would have strong implications for planet formation and dynamical evolution models.
\end{abstract}


\section{Introduction}

NASA's Kepler mission has identified several thousand candidate exoplanet systems \citep{Borucki:2011,Batalha:2012}.  A significant, and important, subset of these systems show multiple candidates orbiting the same star \citep{Steffen:2010}.  From these multiplanet systems we gain important insights into the system architectures, dynamics, and the overall population of planetary systems \citep{Lissauer:2011b,Fabrycky:2012b,Ford:2011a,Fang:2012}.  For example, \citet{Latham:2011} found that short period Jupiter-sized planets do not show additional transiting bodies that have orbital periods within a factor of a few tens.  \citet{Steffen:2012a} subsequently showed that nontransiting planets are not found on orbits near low-order mean-motion resonance (MMR) with hot Jupiter candidates---an observation that can effectively eliminate some models for the formation of hot Jupiter systems.

\citet{Steffen:2012a} made these statements using a bulk search for transit timing variations (TTVs) which are deviations from a constant period that result from gravitational interaction among the various planets within a system \citep{Agol:2005,Holman:2005,Miralda-Escude:2002}.  TTVs have been a boon to exoplanet science from the \kepler\ mission as TTV methods have dynamically confirmed most of the known \kepler\ planets \citep{Holman:2010,Fabrycky:2011,Ford:2011b,Steffen:2012a,Steffen:2013a}, including nontransiting planets \citep{Ballard:2011}, as well as with several unconfirmed systems showing evidence for this signal \citep{Ford:2012a,Steffen:2012a}.

The properties of the \kepler\ multiplanet systems suggest that most planets observable by \kepler\ live in multiplanet systems \citep{Lissauer:2011b,Fabrycky:2012b}.  However, the census of planets in the \kepler\ catalog is incomplete because mutual orbital inclinations among the planets in the systems mean that some planets simply do not transit the host star, and therefore are not detected by the \kepler\ photometry alone.  Nevertheless, \citet{Lissauer:2011b} showed that these mutual orbital inclinations in multiplanet systems are only a few degrees. and that planetary sytems tend to be very flat, as would be expected from planet formation within a gasous protoplanetary disk.  Thus, some systems that have a low multiplicity of observed planets are likely to host nearly coplanar, but unseen sibling planets.

Here we use the orbital period distribution of high multiplicity \kepler\ systems as a means to study lower multiplicity systems.  From this study we can make some statements about where we may expect to find additional nontransiting planets.  The absence of these additional planets would have very significant ramifications regarding the formation and dynamical evolution of the systems in question.  Some models of planet formation suggest that planetary systems are dynamically packed \citep{Laskar:2000,Lissauer:1995,Barnes:2004,Fang:2012a} such that stable orbits in a planetary system are likely to be occupied.  Studies of the architectures and dynamics of \kepler\ systems indicate that virtually all of the \kepler\ multiplanet systems are likely stable and that the ratios of orbital periods are generally not large \citep{Lissauer:2011b,Fabrycky:2012b}.  These observations are consistent with the predictions of dynamical packing.  However, we know from the population of hot Jupiter planets that not all systems share a common dynamical history; and that that relatively small population has motivated new ideas for models of planet formation and dynamical evolution.  The identification of any new population of systems with a unique history would be comparably valuable---such as the long-predicted but newly discovered circumbinary systems \citep{Doyle:2011,Welsh:2012b,Orosz:2012}.  Likewise, the absence of a detectable population of dynamically distinct systems also makes strong statements both about the formation and histories of planetary systems generally and about the very large sample of \kepler\ systems that show only a single planet candidate.

In Section \ref{sechighmult} we investigate some of the properties of the distribution of the period ratios of adjacent planets in systems with four or more observed planet candidates.  Section \ref{seclomult} uses the results from the high multiplicity sample to make and test predictions about what we should see in the low-multiplicity sample where only two planet candidates are seen in each system.  Section \ref{secthree} makes a similar comparison with the three-candidate systems.  We investigate the newly released, preliminary planet candidate list that results from the analysis of \kepler\ data through quarter 8 (Q8) in secction \ref{secq12}.  We make concluding remarks in Section \ref{conclusions}.

\section{High Multiplicity Sample}
\label{sechighmult}

Consider the sample of candidate systems from \citet{Batalha:2012} (hereafter B12) that contain four or more candidates.  We will call this sample the ``high multiplicity'' systems.  \citet{Lissauer:2012} shows that all multiple candidate systems have a high probability of being true planetary systems---a statement that is especially true as the multiplicity (i.e., the number of candidates on a single target) grows.  Thus, we claim that the \kepler\ multiple candidate systems are a sufficiently pure sample of true planetary systems for our study and we will use the terms``planet'', ``planet candidate'', and ``candidate'' are used interchangeably.  The number of high multiplicity systems from B12 is 36 with a total of 154 planets (averaging 4.3 planets per system) however we drop two systems from this sample, one where a planet has a negative period (meaning that only one transit had been seen) and a second where the system is unstable according to the criterion given in \citet{Gladman:1993} assuming Neptune mass planets.  The final high multiplicity sample thus contains 34 planetary systems with 146 planets, giving 114 adjacent planet pairs for study.

\begin{figure}
\includegraphics[width=0.45\textwidth]{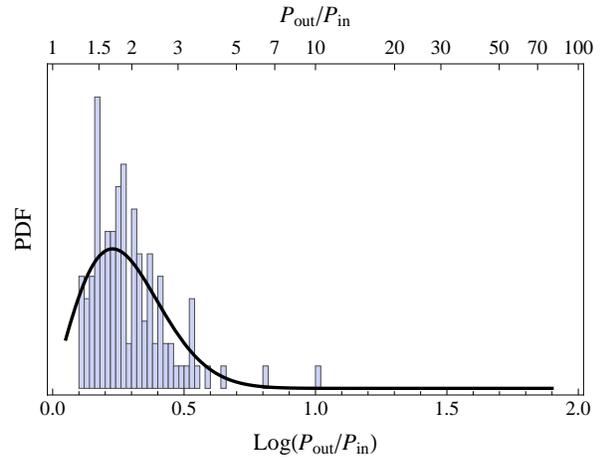}
\caption{Histogram of period ratios for high multiplicity sample.  The binning of this histogram is chosen to show some of the known features near the 3:2 and 2:1 MMRs.}
\label{himulthist}
\end{figure}

\citet{Lissauer:2011b} (hereafter L11) showed that the typical mutual inclination of the orbits in multiplanet systems is a few degrees; multiplanet systems are highly coplanar.  High multiplicity systems likely represent the low mutual-inclination side of that distribution and we use them here as a proxy for multiplanet systems generally.  Figure \ref{himulthist} shows a histogram of the orbital period ratios of adjacent planets in this high multiplicity sample.  In this figure we can see some of the features discussed in L11 and \citet{Fabrycky:2012b}---specifically the excess of planets near the 3:2 MMR and the lack of planets just interior to the 2:1 MMR.  Another feature of potential interest is the marginally significant peak near the 7:2 MMR.  The 7:2 resonance is not normally important dynamically without some modest eccentricity in one or both of the planets in the pair.

The shape of the log of the period ratio distribution is reasonably well approximated by a truncated Rayleigh distribution where the minimum value of the distribution is approximately 0.05 (corresponding to a period ratio of $10^{0.05} = 1.1$ and the Rayleigh parameter is 0.23.  This value of the Rayleigh parameter is found using maximum likelihood estimation on the truncated distribution.  We do not here ascribe any physical significance to this parameterization, only noting that it is generally consistent with the shape of the observed distribution for the high multiplicity sample.  Orbital stability requirements will natually impose some truncation at smaller values as we have done---though the functional form of the distribution at small separations, our approximation notwithstanding, is certainly not a step function.  This distribution tends to over estimate the number of planets in the region roughly between the 2:1 and 4:1 MMRs.  Regardless, for this work neither of the tails of this distribution play a rigorous role and the overestimation of the number of systems at modest period ratios will tend to make later results more conservative.

We expect that most of the planets that lie between the innermost and the outermost planet in the high multiplicity systems are accounted for.  However, some planets will inevitably be absent due to either inclination of the orbit or the smallness of the planet's transit signal (a good example is Kepler-36 \citep{Carter:2012} which had an undetected interior planet that was missed because of its small size and large TTVs).  Missing planets will extend the tail of the period ratio distribution to larger separations.  These large gaps in the observed planetary systems are potentially very interesting since they could contain an additional planet, they could be caused by some physical property of the protoplanetary or planetesimal disk (e.g. the presence of a phase transition in the gas), or they could be a sign of some dynamical event or events in the history of the system.

For observational purposes a simple cut in period ratio would suffice to identify good candidate systems to search for the missing intermediate planets (e.g. all planets where adjacent period ratios are larger than 5:1 or 6:1).  We do, however, wish to make some estimate of the probability that the tail of the distribution is from missing planets rather than simply the natural tail of the distribution.  To do this we iteratively remove data from the long tail of the data and refit a Rayleigh distribution with both tails truncated.  If the data are indeed drawn from this distribution, then the resulting fit to the Rayleigh parameter should be largely independent of the location where the cuts were made.  Outliers, on the other hand, would show up as a decrease in the value of the Rayleigh parameter as the outlier distribution was systematically removed from the analysis---lessening the effects of the tail.  In fact, the value of the Rayleigh parameter does drop from 0.23 to 0.21 as a result of this truncation.  This difference, if measured with high confidence, can cause a nontrivial change to the length of the tail and the height and location of the peak---especially since the period ratio is the exponential of this quantity.

To estimate the statistical significance of this decline in the Rayleigh parameter, we conducted a Monte Carlo simulation where we drew multiple 114-planet samples using a Rayleigh distribution with the large-ratio tail truncated at a variety of values (the small-ratio tail is consistently truncated at a period ratio of 1.1).  The variance in these samples gives us an estimate of the expected statistical fluctuations, which we find to be on the order of 0.1.  This estimate indicates that the widely separated planets cannot be excluded as outliers with high confidence (the decline in the Rayleigh parameter is only $\sim 1.5 \sigma$).  With additional data on these systems, a more sophisticated analysis, or a distribution derived by some other means one might be able to make more definitive statements about the few systems with large gaps in the high multiplicity sample.

Regardless of the statistical significance of the largest gaps, the systems associated with them should prove interesting objects for study.  Mass measurements of the planets in those systems (to see if there is any large change in planet mass near the gaps that might indicate where the formation material accumulated) and searches for additional planets, whether by TTVs or from Radial Velocity (RV) measurements, will yield important information for general planet formation models.  The value of these searches is high regardless of whether additional planets are present or are absent since either result has implications for theoretical models.

\begin{figure}
\includegraphics[width=0.45\textwidth]{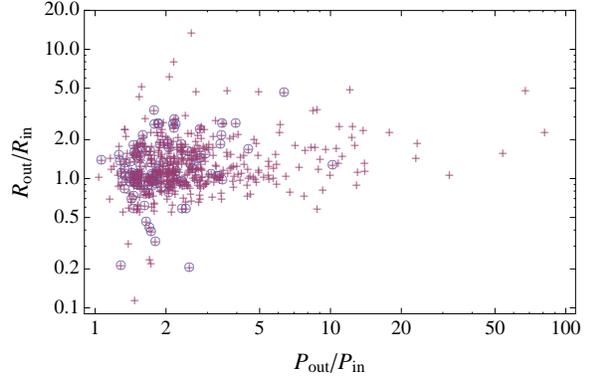}
\caption{Size ratio of adjacent planets as a function of the period ratio of the adjacent planets for the \kepler\ multiplanet systems.  Points corresponding to the high multiplicity sample are highlighted with circles.  We note the increased variance for the most closely spaced pairs.}
\label{persizecorr}
\end{figure}

We note, and show in Figure \ref{persizecorr}, that there is no obvious correlation between the size ratio of adjacent planets and the period ratio of adjacent planets in the multiple candidate systems.  If the largest gaps in the planetary systems corresponded to some important transition in the protoplanetary disk, as opposed to a missing planet, one might expect that the inner and outer planet of the widely separated pair would have dissimilar sizes.  This figure confirms the known obsevation that exterior planets tend to be slightly larger \citep{Ciardi:2013}.  However, there also appears to be larger variance in the size ratio of adjacent planets in more closely spaced orbits than in the more widely separated orbits---both for systems with larger interior and with larger exterior planets.  Some observation bias may be at play here, but the same mechanisms that produced the planets in Kepler-36 \citep{Carter:2012} which have distinctly different densities but nearly the same orbital distance, may also explain this increased variance.  This increase in variance was not observed in \citep{Ciardi:2013} due to the cuts in transit signal-to-noise ratio (SNR) that are applied therein.  Those same cuts are not employed here since our primary concern is with period ratios instead of radius ratios and uncertainties in planet size do not have a significant effect on our results.  Nevertheless, if we consider only those systems with a single transit SNR greater than 10 for all planets in each system, for example, then only $\sim 10$\% of the systems are removed (only one of the high multiplicity systems is removed).  The low SNR systems do not have any observable correlation with period ratio and our results would not be materially affected.

\section{Two-planet Systems}
\label{seclomult}

Now we turn to low multiplicity systems, the 244 systems that have only two transiting planets\footnote{There are 245 systems listed in B12, but one has a planet with a negative (unmeasured) period and is therefore excluded.}.  A reasonable hypothesis about these systems is that they are fundamentally the same as the high multiplicity systems but have slightly larger mutual inclinations.  If this were the case, then the low multiplicity systems should have essentially the same number of planets in them as the high multiplicity sample, but with two or more planets that are not seen simply because they do not transit.  Such systems would still be essentially coplanar as the inclinations need not be large to prevent a planet from transiting.

In principle, it is no more likely that a planet would be missed because of its small size in the low multiplicity sample than it is for the high multiplicity sample.  It is conceivable that many of the low multiplicity sample are in fact high multiplicity systems with all planets transiting but, due to some detection bias that limits the minimum detectable planet size, some planets are missed (e.g., perhaps low multiplicity systems are noisier or more dim).  Several facts, however, do not support this hypothesis.  For example, the median size of the planets in each sample are very similar at 2.1 R$_\oplus$ and 2.2 R$_\oplus$ for the high and low multiplicity samples respectively.  In addition, the average of the first six quarters of Combined Differential Photometric Precision (CDPP) of the high multiplicity sample is slightly worse, with a median of 183 ppm, than the low multiplicity sample, with a median of 128 ppm.  Thus, planets that were missed because of noise would more likely have occurred in the high multiplicity sample.  There remains, however, the caveat that the \kepler\ planet candidate catalog has not yet been uniformly vetted and some more subtle bias could affect the interpretation of these results.

If the low and high multiplicity systems are fundamentally the same, then we should be able to reproduce the period ratio distribution of the low multiplicity systems from the high multiplicity distribution.  We attempt to do this by drawing a sample of planet pairs where the period ratio is either drawn from the best fitting truncated Rayleigh distribution from above, or is generated from two or three realizations of the that distribution being multiplied together (added in log space).  For example, if a low multiplicity were missing the second planet, then the observed period ratio for the first and third planets would be the product of the period ratio of the first and second planets with the ratio of the second and third planets---where both of the ratios would be drawn from the same distribution.  The relative numbers of realizations in this sample with one or two missing planets would be given by the relative probability of a planet to transit $\sim \langle a_\text{in}/a_\text{out} \rangle \simeq 1/1.5$ where $\langle a_\text{in}/a_\text{out} \rangle$ is the mean semimajor axis ratio of the underlying distribution (here we ignore the small contribution that would come from five or six planet systems where three or four planets would be missing).

\begin{figure}
\includegraphics[width=0.45\textwidth]{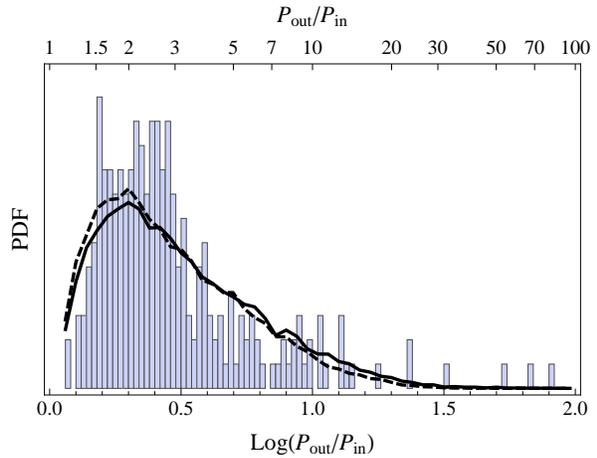}
\caption{Histogram of period ratios for low multiplicity sample as well as two simulated distributions constructed using the Rayleigh distribution that is estimated from the high multiplicity sample (as described in the text).  The solid line corresponds to a Rayleigh parameter of 0.23 and the dashed line has a Rayleigh parameter of 0.21 (which is what one gets when dropping the widely spaced planets---those beyond a ratio of $\sim 6$---from the high multiplicity sample).}
\label{lomulthist}
\end{figure}

We thus construct an approximation to the low multiplicity sample using 10000 realizations from the truncated Rayleigh distribution, 6000 from the product of two realizations, and 4000 from the product of three realizations.  As shown in Figure \ref{lomulthist}, the bulk properties of the low multiplicity distribution can be reproduced in this manner.  However, there are a few notable discrepancies.  The peak is slightly underestimated and the wider separations (between 4:1 and 8:1) are slightly overestimated---as would be expected from the stated over estimate of modest period ratios from the underlying distribution.  Both of these differences can be reduced slightly by using a more narrow model distribution such as what one would get if the largest values in the high multiplicity distribution were considered to be part of an outlier tail.  We note that the period ratio distribution of the inner two planets of the high multiplicity sample is not similar to the low multiplicity sample---indicating that missing planets are needed to explain the overall shape of the low multiplicity distribution.  A similar test, drawing directly from the empirical distribution of the high multiplicity systems yields virtually identical results.

One feature of note in the low multiplicity sample that cannot be readily explained using our simple model is a slight excess of pairs in the region between the 2:1 and 3:1 MMRs.  In general, it is difficult to produce this excess simply by the method described above---that is, by assuming missing planets that are drawn from the high multiplicity distribution.  If we assume that many systems show first and third planets with the second planet missing, then the close-proximity tail of the resulting distribution is far too low.  On the other hand, we assume that most of the low multiplicity distribution is made of true adjacent pairs, then the widely-separated tail of the resulting distribution becomes too low.  Even if we ignore the effects on the two tails, one must still rely on statistical fluctuations to produce the observed excess.

To estimate the significance of this excess under the missing planet model we conduct a Monte Carlo test.  To be conservative, we assume that the probability of seeing the first two planets in a system is equal to missing a second planet while seeing a third---an assumption that favors an excess of longer-period pairs, though not by a large amount.  Of the 244 low-multiplicity \kepler\ candidate pairs, 72 (about 30\%) have period ratios between 2.1 and 2.9.  We drew 10,000 samples of 244 pairs (with 122 that assume no missing planet and 122 that assume the second of three planets is missing) drawn from the truncated Rayleigh distribution shown in Figure \ref{himulthist}---where the period ratio must be greater than 1.1 and the Rayleigh parameter is 0.23.  From these samples only 70 of 10,000 had 72 or more planet pairs in this same region, or less than 1\%.  Thus, there is strong (but not necessarily definitive) evidence for some additional population of planetary systems with period ratios in this region.

If one changes the neighborhood of interest (instead of considering only planet pairs with period ratios between 2.1 and 2.9), then the results of the Monte Carlo simulation give different, though not unexpected results.  In particular, if on considers a narrow neighborhood near the 2:1 MMR, then the probability of seeing the corresponding excess can rise to just under 30\%.  Or, if one considers a narrow neighborhood near the 3:1 MMR, then the probability of seeing the corresponding excess rises to 20\%.  This effect arises primarily because there are relatively fewer candidates observed in the two-planet systems in these narrow windows.  Because of this effect, we believe that averaging over the large neighborhood that we originally state is appropriate for this test.

\begin{figure}
\includegraphics[width=0.45\textwidth]{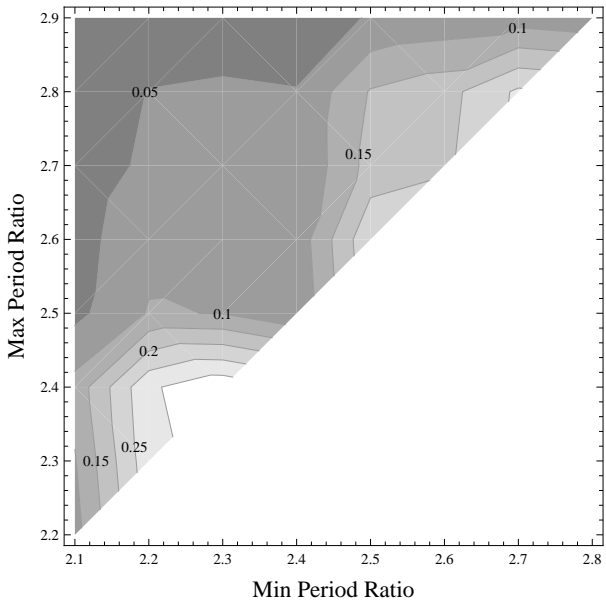}
\caption{The probability, estimated by Monte Carlo simulation, of statistical fluctuations producing the observed excess as a function of the starting and ending locations of the region of interest.  The low probability region in the upper left implies that statistical fluctuations are relatively unlikely to produce the observed excess when considering large regions between the 2:1 and 3:1 MMRs.  The peaks along the diagonal occur because there are fewer observed pairs with these period ratios and one can more easily produce the observed excesses when considering these narrow regions.}
\label{bigmccontour}
\end{figure}

If the underlying Rayleigh distribution is indeed more narrow due to missing planets in the high multiplicity sample, then the probability of producing the observed excess decreases further.  Therefore, the $\sim 0.01$ false alarm probability should be a robust upper bound.  We show, for completeness, in Figure \ref{bigmccontour} the probability of statistical fluctuations producing the excess observed as a function of the values of the start and end of the neighborhood where the excess is measured.  For example, the probability of randomly producing the observed number of planets with period ratios between 2.2 and 2.4 is 0.25 while the probability of randomly producing the number of observed planets with period ratios between 2.4 and 2.8 is 0.05.  Using our model, the expected number of systems in the 2.1 to 2.9 range is approximately 55---as opposed to the 72 observed pairs, implying that perhaps of order 20\% of these systems are anomalous.

A few explanations of this excess are both straightforward and can be tested with further analysis and/or more data.  One possibility is that the observed excess is indeed a statistical fluctuation.  More data from \kepler\ (which we conduct below) or from other exoplanet surveys will test this hypothesis.  A second possibility is that there are, in fact, missing planets in the two planet systems such that the first and third planets transit and have a period ratio between 2 and 3.  While this is expected to be the case some fraction of the time, a challenge for this scenario being the primary explanation for the excess is that it implies that third planets are easier to detect than second planets and that the observed low-multiplicity systems are, in fact, more densly packed than the high multiplicity systems.  A third explanation, which we call the ``resonance'' hypothesis, is that the dynamics of the 2:1, 3:1, 5:2 or other high-order MMRs in this region are strong and planets have become trapped in their orbits due to the presence of these MMRs.  A fourth explanation, which we call the ``collisional'' hypothesis, is that at some point during the formation of the system the orbits of two planets overlapped which eventually led to the collision of those planets.  The consequence of such a collision would be a single remaining planet, on an orbit that was intermediate to the initial pair.  Finally, the observed excess could be due to some other process that occurred either during the formation or subsequent dynamical evolution of the system.  Regardless, further scrutiny of these sytems is justified.

Collisions as a means to explain observations of planetary systems occasionally surfaces (e.g., \citet{Johansen:2012}).  Given the observed tightly packed planets in the \kepler\ high multiplicity sample, it is not unreasonable to believe that some may have gone unstable---with subsequent orbit crossings and collisions.  Should collisions be responsible for the different system architectures, then the distributions of other orbital elements, particularly the eccentricity distribution, may show distinguishing features.  In general collisions tend to reduce eccentricity.  However, this tendency is only true when comparing the eccentricity immediately before and immediately after the collision event and not necessarily when comparing the eccentricities after the collision with the eccentricities from the distant past---before the instability became manifest.

Similarly, there is a strong interplay between eccentricity and mean-motion resonance.  Thus, the resonance hypothesis may also have consequences observable in the orbital element distributions.  Predictions of the form of these distributions, for both the collisional or the resonance hypothesis lies beyond the scope of this work.

\section{Three-planet Systems}
\label{secthree}

While not the primary focus of our study, we now make some comments on the distribution of period ratios in the observed three-planet systems.  There are 83 three planet systems with 166 associated planet pairs that satisfy the stability criterion.  We deliberately excluded three-planet systems from the preceding study so that distinctions between the high multiplicity and low multiplicity systems in the B12 catalog could be more clearly identified.  Nevertheless, the three planet systems also show some interesting features that merit further investigation.

\begin{figure}
\includegraphics[width=0.45\textwidth]{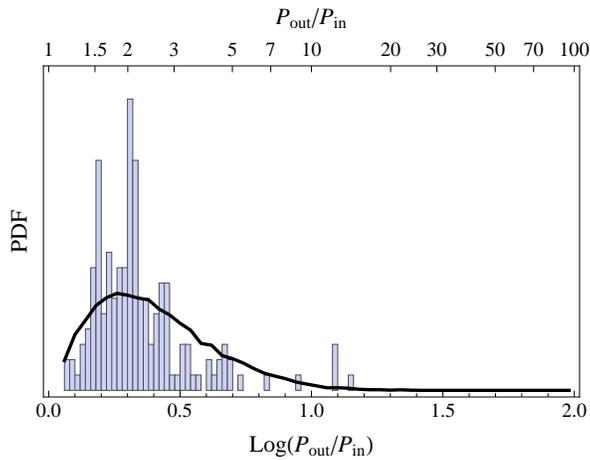}
\caption{Histogram of period ratios for three-planet systems as well as the simulated distributions constructed using the Rayleigh distribution that is estimated from the high multiplicity sample (as described in the text).  The large excess of planets near the 2:1 MMR is not seen in the high multiplicity sample and is more localized than the possible excess seen in the two-planet systems.}
\label{threesystems}
\end{figure}

Figure \ref{threesystems} shows the period ratio distribution for systems with three planet candidates as well as the predicted distribution using a method essentially the same as what was done with the two-planet systems but only assuming a single missing planet.  The most striking feature of the three planet distribution is the large peak surrounding the 2:1 MMR---a peak that is not nearly as prominent in the high multiplicity sample of Figure \ref{himulthist}.  The peak just exterior to the 2:1 and and the small trough that just interior to it (the trough is more readily visible in the high multiplicity sample) has been the subject of additional study \citep{Batygin:2013,Lithwick:2012,Petrovich:2012} and may be generated naturally during the planet formation process or during the subsequent dynamical evolution in the presence of a dissipative force.  In the three-planet systems, the peak near the 2:1 appears to be a true excess and may be caused by orbital migration after the planets have formed \citep{Thommes:2005,Zhou:2005}.  A smaller excess is also visible near the 3:1 MMR that has no counterpart of comparable size in the high multiplicity sample.

Since roughly 25\% of the three planet systems are near the 2:1 MMR, several appear to be near 4:2:1 resonance chains or three-body, Laplace-type resonances.  The observed number of 5 triples where the period ratios of all planets are within 5\% of 4:2:1 is consistent with a random selection from the observed three planet distribution where one would expect 5.2 triples.  Nevertheless, these near-chains contribute materially to the peak near the 2:1 MMR.  It is not clear whether the near 4:2:1 chains are a consequence of or the cause of the excess near 2:1.  Investigating the dynamical properties of these systems will be valuable for identifying their histories.

The presence of the excess planets near the 2:1 MMR in the three-planet systems and the likely excess of planets between the 2:1 and 3:1 MMRs in the two-planet systems may indicate that dynamical interactions play a strong role in determining the number of planets in a system---or at least the number of observed planets.  If the orbits in these systems have modest eccentricities, then the dynamical interactions could excite the inclinations of nearby planets and reduce the number of planets that transit.  Or, strong dynamical interactions may render nearby orbits unstable or preclude planets from forming or migrating into such orbits (though there is no compelling evidence in these data that the third planets in systems with a near-resonant pair have unusually wide separations).  A detailed exploration of these possibilities lies beyond the scope of this work.  As with the features of the two-planet systems, more data and specialized analysis should produce important results that lead to a better understanding of the histories of planetary systems generally.

\section{Results from preliminary Quarter 8 Candidates}
\label{secq12}

In early January 2013 the \kepler\ mission released its current list of KOIs found using data through quarter 8 of operations (Q8).  The public list is preliminary and some changes to it are expected.  Nevertheless, an analysis of this preliminary list is valuable as a means to see if the possible anomalous features in the period ratio distributions seen in the B12 catalog remain when more data are included in the analysis.  We downloaded the KOI list from the NASA Exoplanet Science Institute on January 8, 2013 and conducted tests with those data similar to the tests done for the B12 catalog.  In the Q8 table there are 299 two-candidate systems, 111 three-candidate, and 55 high multiplicity systems (once the stability criterion is invoked).  The basic conclusions from the B12 catalog remain unchanged.  That is, the overall features of the high multiplicity sample remain the same---the location of the peak, the extent of the tail, and the excess near the 3:2 MMR.  The large excess near the 2:1 MMR in the three-planet systems remains prominent as does the small excess just interior to the 3:1 MMR.  We do not show figures of the period ratio distributions from Q8 for the two-planet and three-planet systems as they are not markedly different from the distributions derived from the B12 catalog.

\subsection{Revisited tests of systems with two or three candidates}

With the Q8 data, the number of systems in the low multiplicity sample increased from 244 to 299.  However, there are 90 new two-candidate systems overall.  The difference between these numbers comes from the discovery of new planet candidates in 33 systems (which promotes the systems into the other categories) and two systems that do not appear in the Q8 list.  The discovery of new candidates is not surprising and is consistent with the observation that the period ratio distribution for low-multiplicity systems is largely what one would predict from higher multiplicity systems that simply have undetected planets.  Seven of the low multiplicity systems added two or more planets---joining the high multiplicity sample.  Of these, five showed either additional interior or exterior planet candidates.  The remaining two systems added planets between the two known candidates.  In one of those cases the period ratio of the original pair was 2.22 and the observation of an intermediate planet removes a low multiplicity system from the region of the observed excess.

Many more low multiplicity systems (26 of them) added a single new candidate---joining the three-planet sample.  Of these, a significant fraction were systems with period ratios between the 2:1 and 3:1 MMR where the excess is observed.  Most pairs simply added a new planet either interior to or exterior to the known pair; in some cases the third planet is very widely separated.  However, several two-planet systems (most with period ratios larger than 3:1) added intermediate planets and a large number of the additional planets produced period ratios in the anomalous region near or between the 2:1 and 3:1 MMRs.  That is, the addition of new candidates in two-planet systems from B12 increased the significance of the excesses in the three-planet systems.

\begin{figure}
\includegraphics[width=0.45\textwidth]{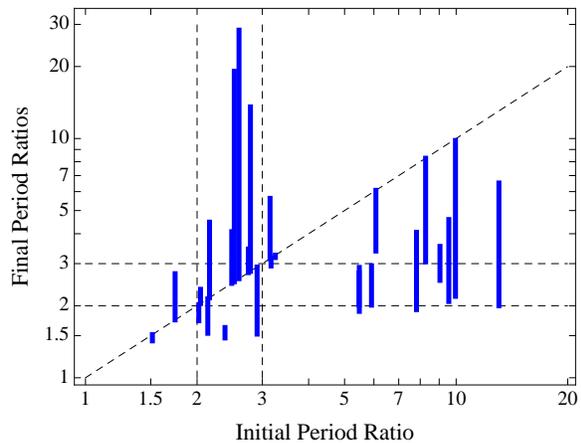}
\caption{A comparison of the initial period ratio of the two-candidate systems to the final period ratios of the three-candidate systems to which they were promoted via the identification of a third planet candidate.  The final period ratios for a given system are joined by a solid vertical line.  The dashed vertical and horizontal lines mark the 2:1 and 3:1 MMRs in both the initial and final samples respectively.  The slanted, dashed line indicates the same initial and final period ratio.  Thus, if a new candidate is discovered interior to or exterior to the initial two-candidate pair, then one of the two ends of the corresponding solid vertical line will be on this slanted line.  If the solid vertical line lies entirely below the slanted line, then an intermediate planet was discovered between the original pair.  No vertical lines lie entirely above the slanted line (which would otherwise indicate that one of the inital periods was an alias of the true period).  Note that several of the initial pairs with large period ratios resulted in triple systems where one or more period ratios are near the 2:1 or 3:1 MMR.  A few of the new KOIs are seen to have very large separations from the initial pair.}
\label{twotothree}
\end{figure}

Figure \ref{twotothree} shows the inital period ratio of each of the two-candidate systems that became three-candidate systems and the two period ratios of the resulting three-candidate systems (shown as a line segment to make the system associations more apparent).  Any line segment that ends on the diagonal line $y=x$ indicates the discovery of a new candidate either interior or exterior to the original pair.  Several of the two-planet systems between the 2:1 and 3:1 MMR (denoted with vertical lines) added planets with relatively large period ratios.  Any line segment that lies entirely below the diagonal line indicates a system where an intermediate planet is discovered.  One can see from the set of line segments that are entirely beneath the diagonal that many planet pairs that initially had large period ratios (beyond 5:1) were actually systems of three planets with many period ratios near the 2:1 and 3:1 MMRs (denoted with horizontal lines) where the anomalous excess was already prominent and is now more so.  There are also a number of systems where the addition of new planets increases the number of planet pairs near the 3:2 MMR.





\begin{figure}
\includegraphics[width=0.45\textwidth]{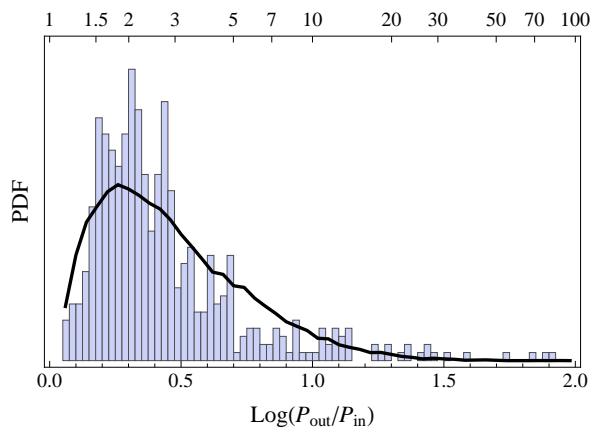}
\caption{The period ratio distribution for the combined two-candidate and three-candidate systems using the Q8 candidate list as well as the predicted distribution from the high multiplicity sample (solid line).  Two significant peaks are visible near the 2:1 and 3:1 MMRs that are not nearly as prominent in the high multiplicity distribution in Figure \ref{himulthist}.  The peak near the 3:2 MMR is visible in all samples.}
\label{twoandthree}
\end{figure}

The result of the additional planets found in the low multiplicity systems from B12 is the removal of a few pairs just inside the 5:2 MMR (and hence a decrease in the significance of the observed excess between the 2:1 and 3:1 MMRs) but a marked increase in the fraction of three planet systems near the 2:1 and 3:1 MMRs.  These facts suggest that the possible excess in the low multiplicity sample between the 2:1 and 3:1 MMR and the observed excess near these MMRs in the three-candidate systems may originate from the same evolutionary processes.  To illustrate this, we show in Figure \ref{twoandthree} a combined period ratio distribution for the two-planet and three-planet systems using the Q8 data as well as the predicted distribution from the high multiplicity sample.  The large peaks near the 2:1 MMR and the 3:1 MMR are not present in the high multiplicity sample (Figure \ref{himulthist}).  Moreover, since the geometricl probability of observing the transits of planet pairs at these larger period ratios the true, debiased period ratio distribution would show these features even more prominently (roughly 25\% larger for the 2:1 MMR and 60\% larger for the 3:1 MMR).  Further study of these systems should shed light on their origin and may offer definitive evidence of a population of exoplanet systems with a unique history.

\begin{figure}
\includegraphics[width=0.45\textwidth]{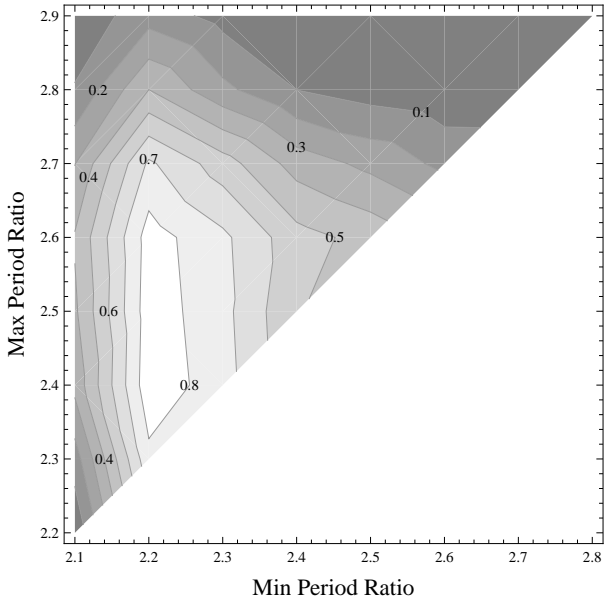}
\caption{The probability, estimated by Monte Carlo simulation, of statistical fluctuations producing the observed excess in the combined two-planet and three-planet period ratio distribution as a function of the starting and ending locations of the region of interest.  The high probability region on the left side is due to the lower number of observed pairs in that region.  The low probability near the 2:1 and 3:1 MMRs implies that statistical fluctuations are unlikely to produce the observed excesss.}
\label{twoandthreecontour}
\end{figure}

A Monte Carlo simulation of the combined two and three planet systems using the period ratios of the high multiplicity systems shows that the observed excess near the 2:1 and 3:1 MMRs are very unlikely to be caused by statistical fluctuations.  Figure \ref{twoandthreecontour} is a contour plot of the probability of statistical fluctuations producing the observed number of systems in a neighborhood as a function of the beginning and ending values that define the neighborhood.  The region where the probability is high corresponds to the region where some two planet systems were removed due to the observation of additional planet candidates in those systems.  Neighborhoods near the 2:1 and 3:1 MMRs have a very low probability that they are statistical in nature.

\subsection{Supplemental comparisons using Quarter 8 Candidates}
\label{secq12sup}

Several additional studies were conducted using both the B12 catalog and the Q8 candidates.  For the sake of brevity we report only the results from the Q8 candidates.  A simple question when comparing the distributions of the high multiplicity and low multiplicity samples is to ask if the period ratio distribution one obtains when considering only the inner two planets of the high multiplicity sample resembles the low multiplicity sample.  In short, the answer is no.  Two-planet systems already have larger period ratios than the high multiplicity systems and removing the outer planets does not decrease this difference---comparing the two distributions using the Kolmogorov-Smirnov (KS) and Anderson-Darling (AD) tests yield $p$ values that are both near $3\times 10^{-4}$.



\begin{figure}
\includegraphics[width=0.45\textwidth]{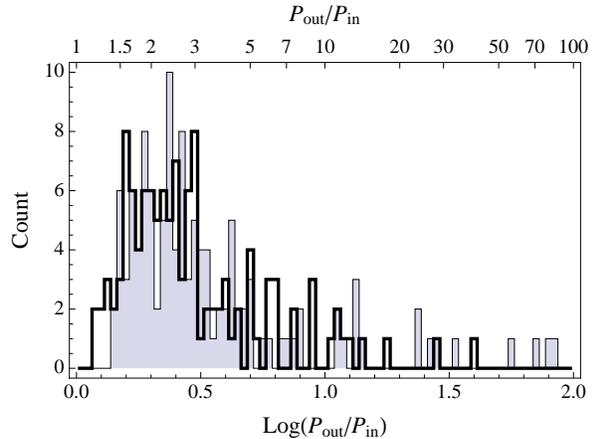}
\caption{Period ratio distribution for the planet pairs in two-planet systems when divided into the 120 smallest systems (40\%), with innermost orbital periods less than 5.69 days (filled), and the 120 largest systems (40\%) with innermost orbital periods greater than 8.47 days (outline).}
\label{twoinandout}
\end{figure}

\begin{figure}
\includegraphics[width=0.45\textwidth]{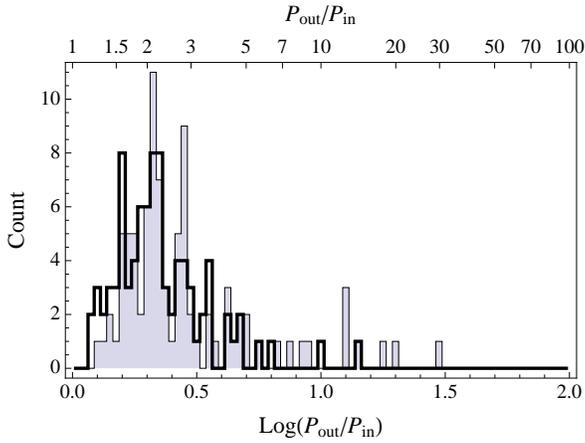}
\caption{Period ratio distribution for the planet pairs in three-planet systems when divided into the 45 smallest systems (40\%), with the innermost orbital period less than 3.89 days (filled), and the 45 largest systems (40\%) with the innermost orbital period more than 6.17 days (outline).}
\label{threeinandout}
\end{figure}

\begin{figure}
\includegraphics[width=0.45\textwidth]{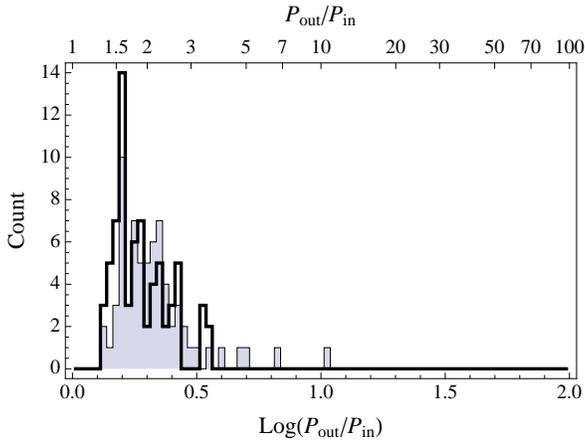}
\caption{Period ratio distribution for the planet pairs in high multiplicity systems when divided into the 22 smallest systems (40\%), with the innermost orbital period less than 3.22 days (filled), and the 22 largest systems (40\%) with the innermost orbital period more than 4.24 days.  As with the previous two figures, this figure is given in counts instead of as a PDF to facilitate the interpretation.  However, for this particular figure, the total number of planets in each subsample is not identical (68 small and 71 large since some systems happen to have a more observed planets).}
\label{largeinandout}
\end{figure}

Another interesting test is to see if there are significant differences in the two and three-planet systems based upon the overall size of the system---that is, innermost orbital period.  Figures \ref{twoinandout}, \ref{threeinandout}, and \ref{largeinandout} compare the period ratio distributions of the smallest 40\% and largest 40\% of the systems (as measured by the orbital period of the inner planet) for two planet, three planet, and high multiplicity systems respectively.  Overall, the distributions for the smaller systems are similar to the distributions for the larger systems, the most significant difference being for the three-planet systems where the Anderson-Darling $p$ value is 0.018 (the $p$ value is 0.043 for high multiplicity systems and 0.13 for two planet systems).  In each case the smaller-sized systems have a larger tail while the larger systems have more planets with small period ratios.  A plausible explanation of this fact may be the orbital decay of systems that are closer to their host stars, where the inner planets become more separated from their counterparts.  Such a process woul both extend the tail of the distribution and reduce the number of systems that are very closely spaced.  A second feature that may prove interesting is that the larger sized systems tend to show a more prominent peak near the 3:2 MMR.  This feature is most obvious in the three-planet systems, though its statistical significance is not large.

\begin{figure}
\includegraphics[width=0.45\textwidth]{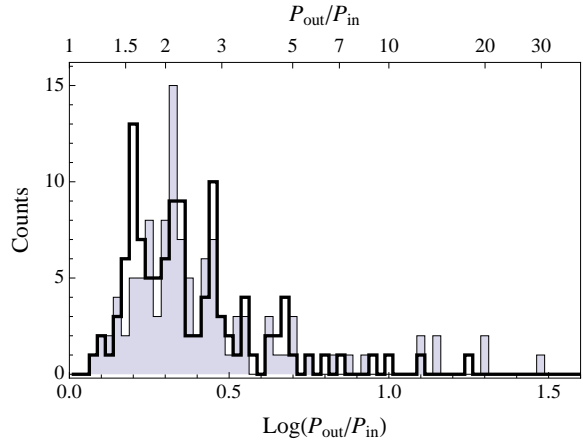}
\caption{A comparison of the period ratio distributions for the inner two planets of the three-planet systems (filled) and the outer pair of planets in the same systems (outline).}
\label{autocomparethree}
\end{figure}

\begin{figure}
\includegraphics[width=0.45\textwidth]{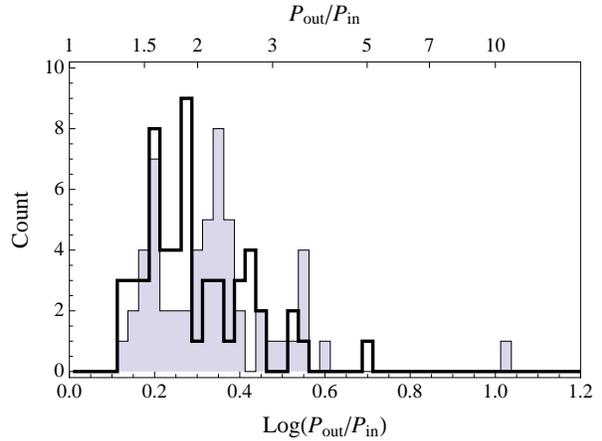}
\caption{A comparison of the period ratio distributions of the inner two planets of the high multiplicity systems (filled) and the third and fourth planets in the same systems (outline).  A comparison with the second and third planets yields similar results.}
\label{autocomparelg}
\end{figure}

A final test that we show is to see if the high multiplicity and three-planet systems are self similar.  That is, does the period ratio distribution for the inner two planets match the distribution for the outer pair (or one of the outer pairs).  Figures \ref{autocomparethree} and \ref{autocomparelg} show the period ratio distributions for the inner and (an) outer planet pair.  For the three planet systems, the inner planet pairs have a prominent peak near the 2:1 MMR while the outer pairs have a more prominent peak near the 3:2 (and perhaps the 3:1).  Globally, however, the distributions are similar and neither the Anderson-Darling test nor the Kolmogorov-Smirnov test indicates a significant difference over the entire distribution.  Considering Poisson fluctuations only in the region near the 3:2 MMR yields a difference that has a $p$ value of about 0.01.

For the high multiplicity systems we consider the specific case of the period ratio distribution of the inner two planets with that obtained for the third and fourth planets (a similar test considering the second and third planets yields similar, and slightly more statistically significant results).  By eye, the difference between the two resulting distributions is striking, even if its significance is less so (the KS test gives a $p$ value of 0.03 and the AD test one of 0.1).  They share a peak near the 3:2 MMR, but with the outer planet pairs have a second peak interior to the 2:1 MMR while the inner planets peak exterior to the 2:1.  Considering poisson fluctuations in the relevant bins yields interesting but not compelling $p$ values near 0.01.  Nevertheless, this feature also exists if you consider the second and third planets instead of the third and fourth.  If this displacement of the peak is a real phenomenon, it too could be explained by the orbital decay of an interior planet.  If planet pairs tend to form interior to the 2:1, the decay of the orbit of the innermost planet (perhaps via tides) would cause the period ratio to grow beyond the 2:1.  The exploration of this possibility is left for elsewhere.

\section{Discussion}
\label{conclusions}

In this paper we explore the differences between the observed period ratio distributions for systems with high multiplicity (four or more planet candidates) and low multiplicity (with two planet candidates) in the \kepler\ catalog published by \citet{Batalha:2012}.  If we make the straightforward assumptions that 1) all of the planets in the high multiplicity systems, out to the outermost one, are accounted for in the \kepler\ data, and 2) that the high multiplicity systems are fundamentally the same as the low multiplicity systems save that two or more of the planets in the low multiplicity systems do not transit, then we observe a few features in the low multiplicity sample that cannot easily be explained using the high multiplicity sample.  The study of the general features of the low multiplicity sample as predicted from the high multiplicity sample, in addition to the study of the anomalous features, has the potential to make important statements about the formation of planetary systems and their dynamical evolution.

One avenue to follow, pertaining to the general properties of the distributions, is that low multiplicity systems where the period ratio of adjacent planets is larger than a few (e.g., $\gtrsim 5$) likely harbor as-yet undetected planets interleaved among the observed ones.  The same may also be true for the high multiplicity systems.  The presence or absence of these intermediate planets would support or refute the model of dynamically full planetary systems---or at least constrain the fraction of systems that are dynamically full.  Large gaps in a planetary system, manifest in the tail of the period ratio distribution, likely indicate either important events or evolutionary processes in that system or important transitions in the structure or composition of the protoplanetary or planetesimal disks.  For example, in our own solar system all of the period ratios of adjacent planets with one exception are well below 3:1---consistent with the high multiplicity distribution.  The exception is the Mars/Jupiter ratio at 6.3, near the possible end of the high multiplicity distribution and where an outlier distribution would start to be observed.  The Mars/Jupiter neighborhood is extremely interesting as it straddles the ice line---a significant phase transition.  Moreover, if Ceres (or the asteroid belt generally) is included in the period ratio distribution, then the gap disappears entirely---replacing a period ratio of 6.3 with a 2.4 and a 2.6.  While the asteroids are not planets in the modern sense, they are a collection of mass located well within the region where our crude model predicts that planets commonly reside.  Distant systems that are roughly similar to our own may have produced a planet there.

A second avenue is to explore the evidence that a small population of planet pairs in the region between the 2:1 and 3:1 MMRs may have some unique history or characteristic.  We do not claim to know the cause of this excess, nor do we claim its existence with high confidence using the B12 catalog---since it is possible at about 1\% of the time ($2.5\sigma$) to produce it in the low multiplicity sample with our simple modeling approach.  We believe, however, that further investigation of the low-multiplicity and three-candidate systems where the planet orbital period ratios are between 2:1 and 3:1 is well motivated.  If it turns out that this excess is simply due to undetected, intermediate planets, then that fact would imply that many planetary systems are more closely packed than the high multiplicity systems suggest (and that there is a modest tendency for an intermediate planet to have either a small size or a relatively large mutual inclination of its orbit).  Intermediate planets may be unlikely in most cases as studies of intermediate orbits (assuming a mass radius relation for the observed candidates) show few that are stable \citep{Fang:2013}.

If it turns out that there are no such additional planets, then one must explain why this excess appears in the systems with only a few planets but not in the high multiplicity systems.  It may be due to strong, ongoing dynamical interactions caused by relatively large eccentricities of the planets in question.  Or, it may be the result of singular events such as planet scattering or collisions.  Additional analysis and more data will help resolve this issue.  While these excess systems in this neighborhood may not be a large population, they may uncover some important insights into the dynamical histories of planetary systems generally.  Such a situation would not be the first time that a relatively small population of planets on unusual orbits gave such insights (e.g., the hot Jupiters).  These systems may have taken longer to detect than the hot Jupiters, but they may prove equally valuable.

\section*{Acknowledgements}
J.H.S. thanks Nick Cowan, Will Farr, Jack Lissauer, Dan Fabrycky, and Darin Ragozzine for insightful conversations leading to this study.  Funding for the \kepler\ mission is provided by NASA's Science Mission Directorate.  We thank the entire Kepler team for the many years of work that is proving so successful.  J.H.S. acknowledges support by NASA under grant NNX08AR04G issued through the Kepler Participating Scientist Program.

\bibliographystyle{plainnat}
\bibliography{multis}

\begin{thebibliography}{36}
\providecommand{\natexlab}[1]{#1}
\providecommand{\url}[1]{\texttt{#1}}
\expandafter\ifx\csname urlstyle\endcsname\relax
  \providecommand{\doi}[1]{doi: #1}\else
  \providecommand{\doi}{doi: \begingroup \urlstyle{rm}\Url}\fi

\bibitem[{Agol} et~al.(2005){Agol}, {Steffen}, {Sari}, and
  {Clarkson}]{Agol:2005}
E.~{Agol}, J.~{Steffen}, R.~{Sari}, and W.~{Clarkson}.
\newblock \emph{\mnras}, 359:\penalty0 567--579, May 2005.

\bibitem[{Ballard} et~al.(2011)]{Ballard:2011}
S.~{Ballard} et~al.
\newblock \emph{\apj}, 743:\penalty0 200, December 2011.

\bibitem[{Barnes} and {Raymond}(2004)]{Barnes:2004}
R.~{Barnes} and S.~N. {Raymond}.
\newblock \emph{\apj}, 617:\penalty0 569--574, December 2004.

\bibitem[{Batalha} et~al.(2012)]{Batalha:2012}
N.~M. {Batalha} et~al.
\newblock \emph{ArXiv e-prints}, February 2012.

\bibitem[{Batygin} and {Morbidelli}(2013)]{Batygin:2013}
K.~{Batygin} and A.~{Morbidelli}.
\newblock \emph{\aj}, 145:\penalty0 1, January 2013.

\bibitem[{Borucki} et~al.(2011)]{Borucki:2011}
W.~J. {Borucki} et~al.
\newblock \emph{\apj}, 736:\penalty0 19--+, July 2011.

\bibitem[{Carter} et~al.(2012){Carter}, {Agol}, et~al.]{Carter:2012}
J.~A. {Carter}, E.~{Agol}, et~al.
\newblock \emph{ArXiv e-prints:1206.4718}, June 2012.

\bibitem[{Ciardi} et~al.(2012){Ciardi}, {Fabrycky}, {Ford}, {Gautier},
  {Howell}, {Lissauer}, {Ragozzine}, and {Rowe}]{Ciardi:2013}
D.~R. {Ciardi}, D.~C. {Fabrycky}, E.~B. {Ford}, T.~N. {Gautier}, III, S.~B.
  {Howell}, J.~J. {Lissauer}, D.~{Ragozzine}, and J.~F. {Rowe}.
\newblock \emph{ArXiv e-prints}, December 2012.

\bibitem[{Doyle} et~al.(2011)]{Doyle:2011}
L.~R. {Doyle} et~al.
\newblock \emph{Science}, 333:\penalty0 1602--, September 2011.

\bibitem[{Fabrycky} et~al.(2011){Fabrycky}, {Ford}, {Steffen},
  et~al.]{Fabrycky:2011}
D.~C. {Fabrycky}, E.~B. {Ford}, J.-H. {Steffen}, et~al.
\newblock \emph{\apj\ Submitted}, 2011.

\bibitem[{Fabrycky} et~al.(2012)]{Fabrycky:2012b}
D.~C. {Fabrycky} et~al.
\newblock \emph{ArXiv e-prints:1202.6328}, February 2012.

\bibitem[{Fang} and {Margot}(2012{\natexlab{a}})]{Fang:2012}
J.~{Fang} and J.-L. {Margot}.
\newblock \emph{ArXiv e-prints:1207.5250}, July 2012{\natexlab{a}}.

\bibitem[{Fang} and {Margot}(2012{\natexlab{b}})]{Fang:2012a}
J.~{Fang} and J.-L. {Margot}.
\newblock \emph{\apj}, 751:\penalty0 23, May 2012{\natexlab{b}}.

\bibitem[{Fang} and {Margot}(2013)]{Fang:2013}
J.~{Fang} and J.-L. {Margot}.
\newblock \emph{ArXiv e-prints:1302.7190}, February 2013.

\bibitem[{Ford} et~al.(2011{\natexlab{a}}){Ford}, {Fabrycky}, {Steffen},
  et~al.]{Ford:2011b}
E.~B. {Ford}, D.~C. {Fabrycky}, J.-H. {Steffen}, et~al.
\newblock \emph{\apj\ Submitted}, 2011{\natexlab{a}}.

\bibitem[{Ford} et~al.(2012){Ford}, {Fabrycky}, {Steffen}, et~al.]{Ford:2012a}
E.~B. {Ford}, D.~C. {Fabrycky}, J.~H. {Steffen}, et~al.
\newblock \emph{\apj}, 750:\penalty0 113, May 2012.

\bibitem[{Ford} et~al.(2011{\natexlab{b}})]{Ford:2011a}
E.~B. {Ford} et~al.
\newblock \emph{\apjs}, 197:\penalty0 2--+, November 2011{\natexlab{b}}.

\bibitem[{Gladman}(1993)]{Gladman:1993}
B.~{Gladman}.
\newblock \emph{\icarus}, 106:\penalty0 247, November 1993.

\bibitem[{Holman} and {Murray}(2005)]{Holman:2005}
M.~J. {Holman} and N.~W. {Murray}.
\newblock \emph{Science}, 307:\penalty0 1288--1291, February 2005.

\bibitem[{Holman} et~al.(2010)]{Holman:2010}
M.~J. {Holman} et~al.
\newblock \emph{Science}, 330:\penalty0 51--, October 2010.

\bibitem[{Johansen} et~al.(2012){Johansen}, {Davies}, {Church}, and
  {Holmelin}]{Johansen:2012}
A.~{Johansen}, M.~B. {Davies}, R.~P. {Church}, and V.~{Holmelin}.
\newblock \emph{\apj}, 758:\penalty0 39, October 2012.

\bibitem[{Laskar}(2000)]{Laskar:2000}
J.~{Laskar}.
\newblock \emph{Physical Review Letters}, 84:\penalty0 3240--3243, April 2000.

\bibitem[{Latham} et~al.(2011)]{Latham:2011}
D.~W. {Latham} et~al.
\newblock \emph{\apjl}, 732:\penalty0 L24+, May 2011.

\bibitem[{Lissauer}(1995)]{Lissauer:1995}
J.~J. {Lissauer}.
\newblock {Urey prize lecture: On the diversity of plausible planetary
  systems}.
\newblock \emph{\icarus}, 114:\penalty0 217--236, April 1995.

\bibitem[{Lissauer} et~al.(2011)]{Lissauer:2011b}
J.~J. {Lissauer} et~al.
\newblock \emph{\apjs}, 197:\penalty0 8, November 2011.

\bibitem[{Lissauer} et~al.(2012)]{Lissauer:2012}
J.~J. {Lissauer} et~al.
\newblock \emph{\apj}, 750:\penalty0 112, May 2012.

\bibitem[{Lithwick} and {Wu}(2012)]{Lithwick:2012}
Y.~{Lithwick} and Y.~{Wu}.
\newblock \emph{\apjl}, 756:\penalty0 L11, September 2012.

\bibitem[{Miralda-Escud{\'e}}(2002)]{Miralda-Escude:2002}
J.~{Miralda-Escud{\'e}}.
\newblock \emph{\apj}, 564:\penalty0 1019--1023, January 2002.

\bibitem[{Orosz} et~al.(2012)]{Orosz:2012}
J.~A. {Orosz} et~al.
\newblock \emph{\apj}, 758:\penalty0 87, October 2012.

\bibitem[{Petrovich} et~al.(2012){Petrovich}, {Malhotra}, and
  {Tremaine}]{Petrovich:2012}
C.~{Petrovich}, R.~{Malhotra}, and S.~{Tremaine}.
\newblock \emph{ArXiv e-prints}, November 2012.

\bibitem[{Steffen} et~al.(2012){Steffen}, {Fabrycky}, {Ford},
  et~al.]{Steffen:2012a}
J.~H. {Steffen}, D.~C. {Fabrycky}, E.~B. {Ford}, et~al.
\newblock \emph{\mnras}, 421:\penalty0 2342--2354, April 2012.

\bibitem[{Steffen} et~al.(2010)]{Steffen:2010}
J.~H. {Steffen} et~al.
\newblock \emph{\apj}, 725:\penalty0 1226--1241, December 2010.

\bibitem[{Steffen} et~al.(2013)]{Steffen:2013a}
J.~H. {Steffen} et~al.
\newblock \emph{\mnras}, 428:\penalty0 1077--1087, January 2013.

\bibitem[{Thommes}(2005)]{Thommes:2005}
E.~W. {Thommes}.
\newblock \emph{\apj}, 626:\penalty0 1033--1044, June 2005.

\bibitem[{Welsh} et~al.(2012)]{Welsh:2012b}
W.~F. {Welsh} et~al.
\newblock \emph{\nat}, 481:\penalty0 475--479, January 2012.

\bibitem[{Zhou} et~al.(2005){Zhou}, {Aarseth}, {Lin}, and
  {Nagasawa}]{Zhou:2005}
J.-L. {Zhou}, S.~J. {Aarseth}, D.~N.~C. {Lin}, and M.~{Nagasawa}.
\newblock \emph{\apjl}, 631:\penalty0 L85--L88, September 2005.

\end{thebibliography}

\bsp

\label{lastpage}

\end{document}